\documentclass[letter,12pt]{article}

\usepackage{jheppub}

\usepackage{graphicx,amssymb,amsmath,color}
\usepackage{hyperref}

\def\beq{\begin{equation}}
\def\eeq{\end{equation}}

\def\bea{\begin{eqnarray}}
\def\eea{\end{eqnarray}}
\def\bei{\begin{itemize}}
\def\eei{\end{itemize}}
\def\bmat{\begin{matrix}}
\def\emat{\end{matrix}}
\def\={\,=\,}
\def\+{\,+\,}
\def\-{\,-\,}

\newcommand{\Fig}[1]{Fig.~\ref{#1}}
\newcommand{\Eq}[1]{Eq.~(\ref{#1})}
\newcommand{\Sec}[1]{Sec.~\ref{#1}}

\title{Very Degenerate Higgsino Dark Matter}

\author[a]{Eung Jin Chun,}
\emailAdd{ejchun@kias.re.kr}
\affiliation[a]{Korea Institute for Advanced Study, Seoul 130-722, Korea}

\author[b,c]{Sunghoon Jung,}
\emailAdd{shjung@slac.stanford.edu}
\affiliation[b]{SLAC National Accelerator Laboratory, Menlo Park, CA 94025, USA}
\affiliation[c]{Kavli Institute for Theoretical Physics, Santa Barbara, CA 93106, USA}

\author[d]{Jong-Chul Park}
\emailAdd{jcpark@cnu.ac.kr}
\affiliation[d]{Department of Physics, Chungnam National University, Daejeon 34134, Korea}

\abstract{
We present a study of the Very Degenerate Higgsino Dark Matter (DM), whose mass splitting between the lightest neutral and charged components is ${\cal O}(1)$ MeV, much smaller than radiative splitting of 355 MeV. The scenario is realized in the minimal supersymmetric standard model by small gaugino mixing. In contrast to the pure Higgsino DM with the radiative splitting only, various observable signatures with distinct features are induced. First of all, the very small mass splitting makes (a) sizable Sommerfeld enhancement and Ramsauer-Townsend (RT) suppression relevant to $\sim$1 TeV Higgsino DM, and (b) Sommerfeld-Ramsauer-Townsend effect saturate at lower velocities $v/c \lesssim 10^{-3}$. As a result, annihilation signals can be large enough to be observed from the galactic center and/or dwarf galaxies, while relative signal sizes can vary depending on the location of Sommerfeld peaks and RT dips. In addition, at collider experiments, stable chargino signature can be searched for to probe the model in the future. DM direct detection signal, however, depends on the Wino mass; even no detectable signal can be induced if the Wino is heavier than about 10 TeV.
}

\preprint{SLAC-PUB-16697, NSF-KITP-16-092}

\begin{document}
\maketitle

\section{Introduction}

The pure Higgsino (with the electroweak-radiative mass splitting $\Delta m=355$ MeV between its lightest neutral and charged components) is an attractive candidate of thermal dark matter (DM) for its mass around 1 TeV~\cite{ArkaniHamed:2006mb}.
As null results at Large Hadron Collider (LHC) experiments push supersymmetry (SUSY) to TeV scale, such Higgsino as the lightest supersymmetric particle (LSP) has recently become an important target for future collider~\cite{Low:2014cba, Acharya:2014pua, Gori:2014oua, Barducci:2015ffa, Badziak:2015qca, Bramante:2015una} and DM search experiments~\cite{Barducci:2015ffa, Badziak:2015qca, Bramante:2015una, Cirelli:2007xd, Chun:2012yt, Fan:2013faa, Chun:2015mka}.
\emph{A priori}, the Higgsino mass $\mu$ and gaugino masses $M_1, M_2$ for the Bino and Wino are not related; thus, the pure Higgsino scenario with much heavier gauginos is possible and natural by considering two distinct Peccei-Quinn and R symmetric limits.

It is, however, difficult to test the pure Higgsino LSP up to 1--2 TeV at collider experiments (including future 100 TeV options) and dark matter detections.
Standard collider searches of jet plus missing energy are insensitive because of the small mass splitting of 355 MeV~\cite{Low:2014cba, Acharya:2014pua, Gori:2014oua}; but the splitting is large enough for charginos to decay promptly at collider so that disappearing track and stable chargino searches are not sensitive~\cite{Low:2014cba, Thomas:1998wy}.
The purity of the Higgsino states suppresses DM direct detection signals.
DM indirect detection signals are not large enough because of relatively weak interactions and negligible Sommerfeld enhancements~\cite{Hisano:2003ec, Hisano:2004ds, Cirelli:2007xd, Chun:2012yt, Fan:2013faa, Chun:2015mka}.
In contrast, the pure Wino DM with the radiative mass splitting of 164 MeV, another thermal DM candidate for its mass $\sim$ 3 TeV, provides several ways to test: monojet plus missing energy due to more efficient recoil and larger cross-section~\cite{Low:2014cba, Gori:2014oua, Cirelli:2014dsa, Bhattacherjee:2012ed}, disappearing track due to longer-lived charged Wino~\cite{Low:2014cba, Cirelli:2014dsa, Bhattacherjee:2012ed}, and indirect detection due to somewhat stronger interaction and larger enhancement~\cite{Hisano:2004ds, Cirelli:2007xd, Chun:2012yt, Fan:2013faa, Cohen:2013ama, Chun:2015mka}.
One of the key features of the Wino DM affecting all of these signals is the smaller mass splitting.

It has been noticed that the non-perturbative effects can be sizable for the heavy electroweak dark matter annihilation, leading to not only the Sommerfeld enhancement~\cite{Hisano:2003ec, Hisano:2004ds} but also
the Ramsauer-Townsend (RT) suppression~\cite{Chun:2012yt, Chun:2015mka, Cirelli:2015bda, Garcia-Cely:2015dda} that become more evident for smaller mass splitting (or equivalently heavier DM) and higher multiplets (or stronger electroweak interactions)~\cite{Chun:2012yt, Chun:2015mka}.
The Higgsino-gaugino system, consisting of the weak singlet, doublet and triplet, with variable mass splitting provides a natural framework realizing drastic Sommerfeld-Ramsauer-Townsend (SRT) effects in dark matter annihilation.

This motivates us to investigate a possibility of a very degenerate Higgsino DM whose mass splitting is much smaller than the electroweak-induced 355 MeV, realized in the limit of $\mu \ll M_{1, 2}$ admitting slight gaugino mixtures.
The Higgsino is more susceptible to nearby gauginos than the gaugino is to others as heavier gaugino effects on the Higgsino decouple less quickly:
their effects are captured by dimension-5 operators, while effects on the gaugino DM is captured by dimension-6 operators~\cite{Fan:2013faa}.
Thus, it leads to a plausible situation that heavier gauginos are almost decoupled leaving some traces only in the Higgsino DM sector in spite of a large hierarchy between them.
The Very Degenerate Higgsino DM turns out to produce distinct features in indirect detection signals from the galactic center (GC) and dwarf spheroidal satellite galaxies (DG), which can be observed in the near future.

This paper is organized as follows.
In Section 2, we look for the Higgsino-gaugino parameter space realizing the Very Degenerate Higgsino LSP.
In Section 3, indirect signals of DM annihilation are studied to feature the SRT effect,
which leads to distinct predictions for the GC and DG.
In Section 4, we consider other constraints from direct detection, collider searches, and cosmology.
We finally conclude in Section 5.

\section{Very Degenerate Higgsino DM}

We discuss the SUSY parameter space of the Very Degenerate Higgsino DM, which involves the Higgsino mass parameter $\mu$, the Bino and Wino masses $M_{1,2}$, the ratio of the Higgs vacuum expectation values $t_\beta\equiv \tan\beta=v_u/v_d$, the weak mixing angle given by $s_W\equiv \sin\theta_W \approx 0.23$, and the $W$ gauge boson mass $m_W$.
We assume the limit $|M_1 \pm M_2|, |M_2 \pm \mu|, |\mu \pm M_1| \gg m_W$.
We keep the signs of mass eigenvalues and make eigenvectors real.
Later on, we will assume $M_2, \mu >0$ and $M_1<0$ for the Very Degenerate Higgsino DM, but we will be agnostic about how such signs can be obtained.

Higgsino mass eigenvalues at tree-level are~\cite{Drees:1996pk},
\bea
m_{\chi^+} &\simeq& |\mu| - {\rm sgn}(\mu M_2) \frac{m_W^2}{|M_2|}  s_{2\beta} \, > \, 0\,, \\
m_{\chi^0_{S, A}} &\simeq& \mp \mu - \frac{m_W^2}{2 M_2}(1 \mp s_{2\beta}) \, \epsilon_K\,, \qquad \epsilon_K \equiv \left( 1 + \frac{M_2}{M_1} t_W^2 \right)\,,
\eea
where $s_{2\beta} = \sin 2\beta$ and so on.
The subscripts $S, A$ imply that the mass eigenstates are $\chi^0_{S, A} \sim (\widetilde{H}_d^0 \pm \widetilde{H}_u^0)/\sqrt{2}$.
Which of $\chi^0_{S}$ or $\chi^0_A$ is the LSP depends on the relative sign of $\mu$ and $\epsilon_K M_2$: the $\chi^0_A$ is the LSP if the relative sign is positive, and vice versa.
Expressing both possibilities, we write the LSP mass as
\beq
m_{\chi^0_1} \, \simeq \, {\rm sgn}(\epsilon_K M_2) \left(  |\mu | - \frac{m_W^2}{2 |M_2|} \Big( 1+ {\rm sgn}(\mu \epsilon_K M_2) \cdot s_{2\beta} \Big) \, | \epsilon_K | \right)\,.
\eeq
Higgsino mass splitting at tree-level is then
\bea
\Delta m_{\rm tree} &\equiv& m_{\chi^+} - | m_{\chi^0_1} |  \nonumber \\
&\simeq& - {\rm sgn}(\mu M_2) \frac{m_W^2}{|M_2|} s_{2\beta} \+ \frac{m_W^2}{2|M_2|} \Big( 1 + {\rm sgn}(\mu \epsilon_K M_2) \cdot s_{2\beta} \Big) \, |\epsilon_K|\,.
\label{eq:deltam} \eea
The physical mass splitting is $\Delta m = \Delta m_{\rm tree} + \Delta m_{\rm loop}$, where the model-independent electroweak loop corrections give $\Delta m_{\rm loop} \approx 355$ MeV for the Higgsino~\cite{Thomas:1998wy}.

Notably, the $\Delta m_{\rm tree}$ can be negative, so that the resulting physical mass splitting $\Delta m$ can be smaller than the $\Delta m_{\rm loop}$.\footnote{The negative $\Delta m_{\rm tree}$ has been used in exotic collider phenomenology of Higgsinos~\cite{Kribs:2008hq, Jung:2015boa}.}
From the above approximations, we find that one way to obtain negative $\Delta m_{\rm tree}$ is to satisfy the following conditions:
\begin{itemize}
\item sign($\mu M_2)>0$ because only the first term in \Eq{eq:deltam} can be negative.
Assuming $\mu, \, M_2 >0$ from now on, we rewrite
\bea
\Delta m_{\rm tree} &\simeq& \left\{ \bmat  - \frac{m_W^2}{M_2} \, \left( \, s_{2\beta} \- \frac{\epsilon_K}{2} \Big( 1 + s_{2\beta} \Big) \right) && {\rm for} \quad \epsilon_K>0 \\
- \frac{m_W^2}{M_2} \, \left( \, s_{2\beta} \+ \frac{\epsilon_K}{2} \Big( 1 - s_{2\beta} \Big) \right) && {\rm for} \quad \epsilon_K<0  \emat \right.\,.
\label{eq:deltam2} \eea
Thus, $\Delta m_{\rm tree}<0$ if masses satisfy
\beq
-\frac{2 s_{2\beta}}{1-s_{2\beta}} \, \lesssim \, \epsilon_K \, \lesssim \, \frac{2 s_{2\beta}}{1+s_{2\beta}} \leq 1\,.
\label{eq:cond-epk} \eeq
\item $M_1 <0$ is preferred so that $\epsilon_K<1$.
\item Small $t_\beta$ is preferred.
\item In the limit of $M_2 \to \infty$ or $M_1 \to \infty$, no solutions exist.
\end{itemize}
We apply this set of approximate conditions to our full numerical calculation to narrow down solution finding procedure.

\begin{figure}[t] \centering
\includegraphics[width=0.65\textwidth]{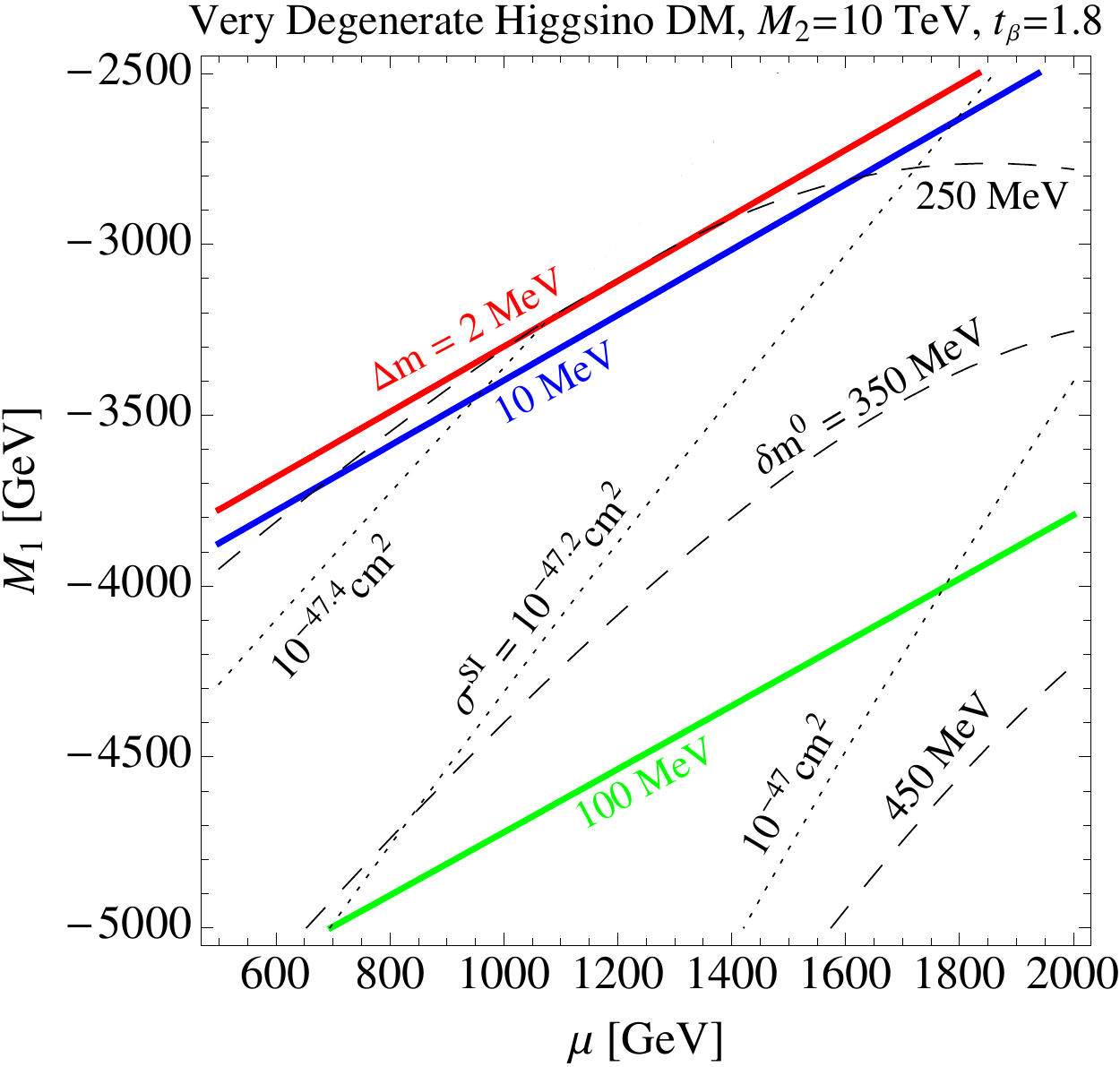}
\caption{Very Degenerate Higgsino DM parameter space with $M_2$=10 TeV and $t_\beta=1.8$.
Contours of the mass splitting $\Delta m = m_{\chi^+} - |m_{\chi_1^0}|$ (solid), $\delta m^0 = |m_{\chi_2^0}| - |m_{\chi^0_1}|$ (dashed), and spin-independent direct detection rate $\sigma^{\rm SI}$ (dotted; see \Sec{sec:dd}) are shown. We consider the two benchmark models along the $\Delta m=$ 2, 10 MeV contours throughout.
}
\label{fig:mu-m1-masseigen}
\end{figure}

In \Fig{fig:mu-m1-masseigen}, we show one set of numerical solutions of $\Delta m_{\rm tree} <0$ for the range of $\mu \leq 2$ TeV and $-2.5 \geq M_1 \geq -5$ TeV with fixed $M_2 = 10$ TeV and $t_\beta = 1.8$.
In most of the parameter space shown, $\Delta m$ is smaller than the radiative mass splitting of 355 MeV.
Although approximate equations above do not depend on $\mu$, the full numerical solution does a bit.
We will consider two benchmark cases of $\Delta m=$2, 10 MeV in this parameter space throughout.
Later, we will also comment on the case with smaller $M_2 = 5$ TeV. 
The solutions for $\Delta m=$2, 10 MeV and our most discussions do not strongly depend on the value of $M_2$, but direct detection signal does as will be discussed.
The neutralino mass splitting, $\delta m^0 \equiv |m_{\chi_2}^0| - |m_{\chi_1}^0|$, is somewhat larger $\sim {\cal O}(100)$ MeV, and it also does not strongly affect our discussion.

\section{Indirect Detection of Annihilation Signals}

Non-perturbative effects in DM pair annihilation can lead to Sommerfeld enhancement~\cite{Hisano:2003ec, Hisano:2004ds} or Ramsauer-Townsend suppression~\cite{Chun:2012yt, Chun:2015mka}.
The pure Higgsino DM with $\mu \sim 1$ TeV and $\Delta m\approx 355$ MeV does not experience large SRT effects.
Only Higgsinos as heavy as $\sim 7$ TeV can experience sizable effects, but they are too heavy to be relevant to collider experiments.
On the other hand, 1--3 TeV pure Wino DM with $\Delta m \approx 164$ MeV experiences much larger SRT effects with a resonance appearing at around 2.4 TeV~\cite{Hisano:2003ec, Hisano:2004ds, Cirelli:2007xd, Chun:2012yt, Chun:2015mka, Fan:2013faa, Cohen:2013ama}. Since the SRT effects on the pure Wino DM saturate at relatively high velocity $v/c \sim 10^{-2}$, Wino annihilation cross-sections at various astronomical sites with different velocity dispersions are same.

We will discuss that the very small splitting of the Higgsino DM can make the relevant Higgsino mass scale down to $\sim 1$ TeV and allow different annihilation cross-sections at various astronomical sites, postponing the saturation to lower velocities.
Furthermore, there can appear not only Sommerfeld enhancements but also RT suppressions.

\subsection{SRT Effects with Very Small Mass Splitting}

We focus on today's DM annihilation cross-sections into $WW, ZZ, \gamma\gamma, Z\gamma$ channels.
Thus, we do not consider co-annihilation channels.
Pair annihilations with SRT effects can proceed via various intermediate two-body states with the same charge $Q=0$ and spin $S=0, 1$ as those of the initial LSP pair, which are exchanged by photons and on/off-shell $W, Z$ gauge bosons.
We take into account all two-body states formed among Higgsino states; in addition, we add heavier gauginos if their masses are within 10 GeV of the Higgsino in order to accommodate non-zero effects from them, but this rarely happens in our study.
We follow a general formalism developed for SUSY in Ref.~\cite{Beneke:2012tg, Hellmann:2013jxa, Beneke:2014gja, Beneke:2014hja} to calculate absorptive Wilson coefficients and non-relativistic potentials between various two-body states, and we numerically solve resulting Schr$\ddot{\rm o}$dinger equations to obtain SRT effects.

\begin{figure}[t] \centering
\includegraphics[width=0.489\textwidth]{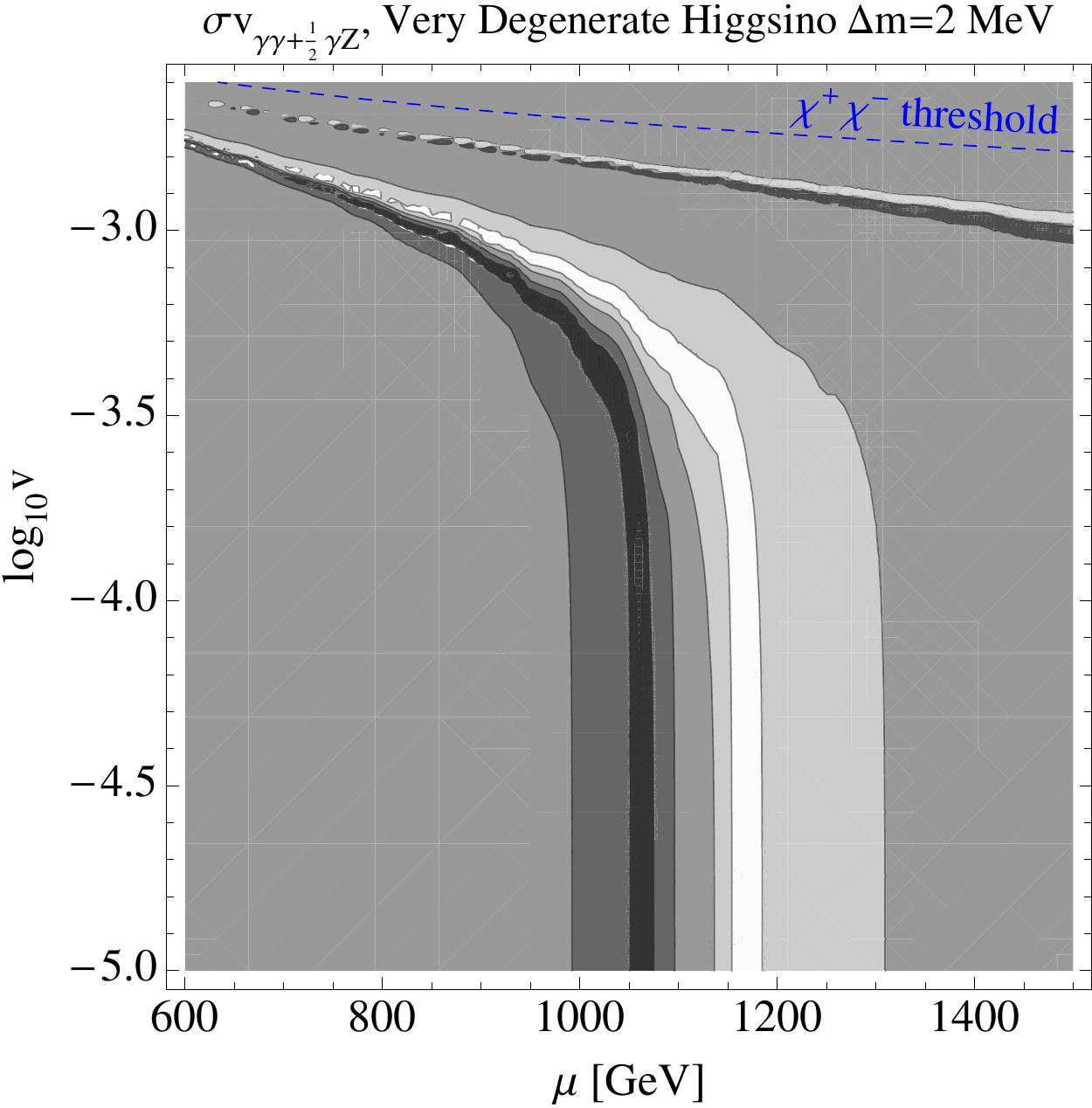}
\includegraphics[width=0.502\textwidth]{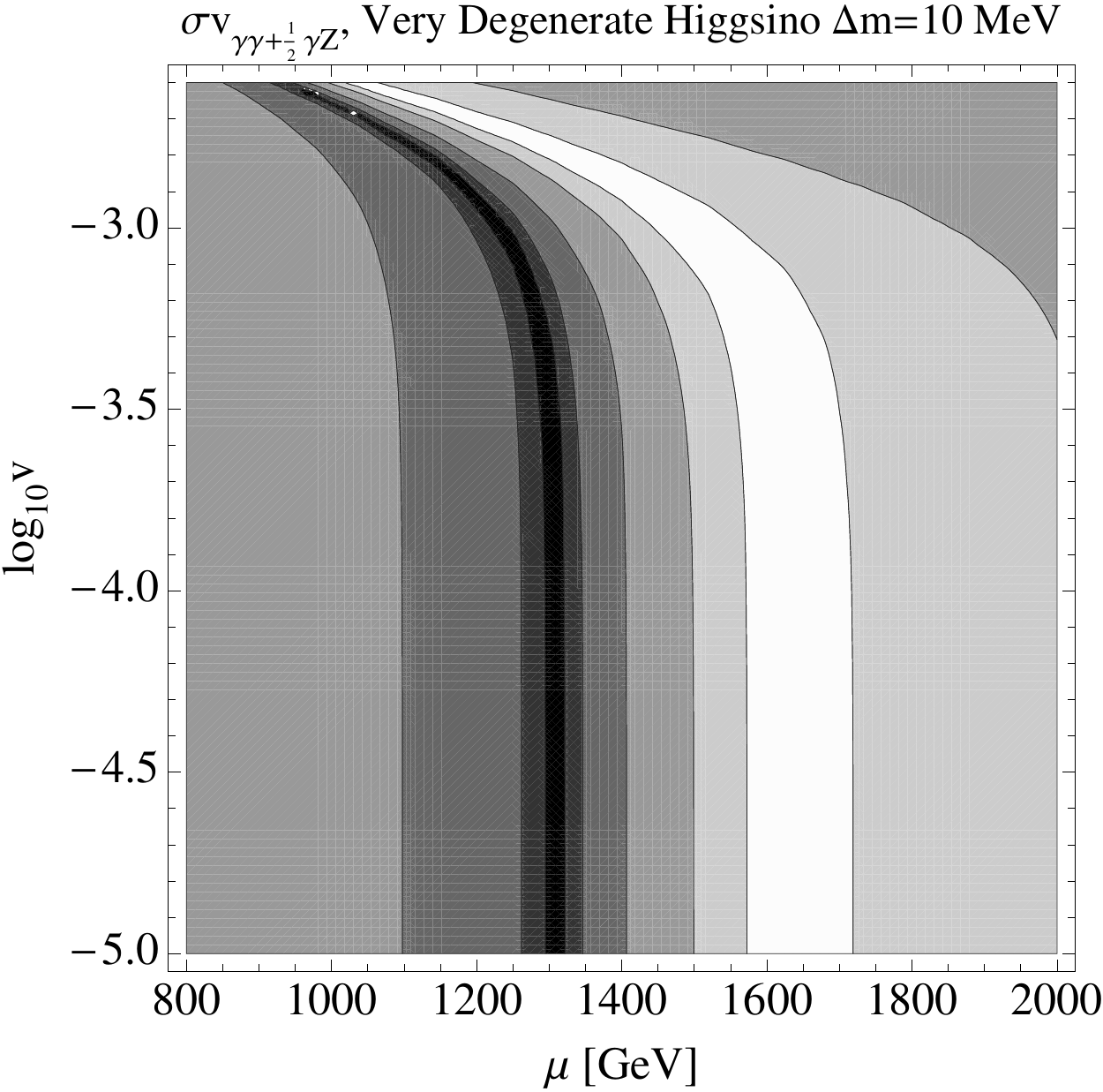}
\includegraphics[width=0.50\textwidth]{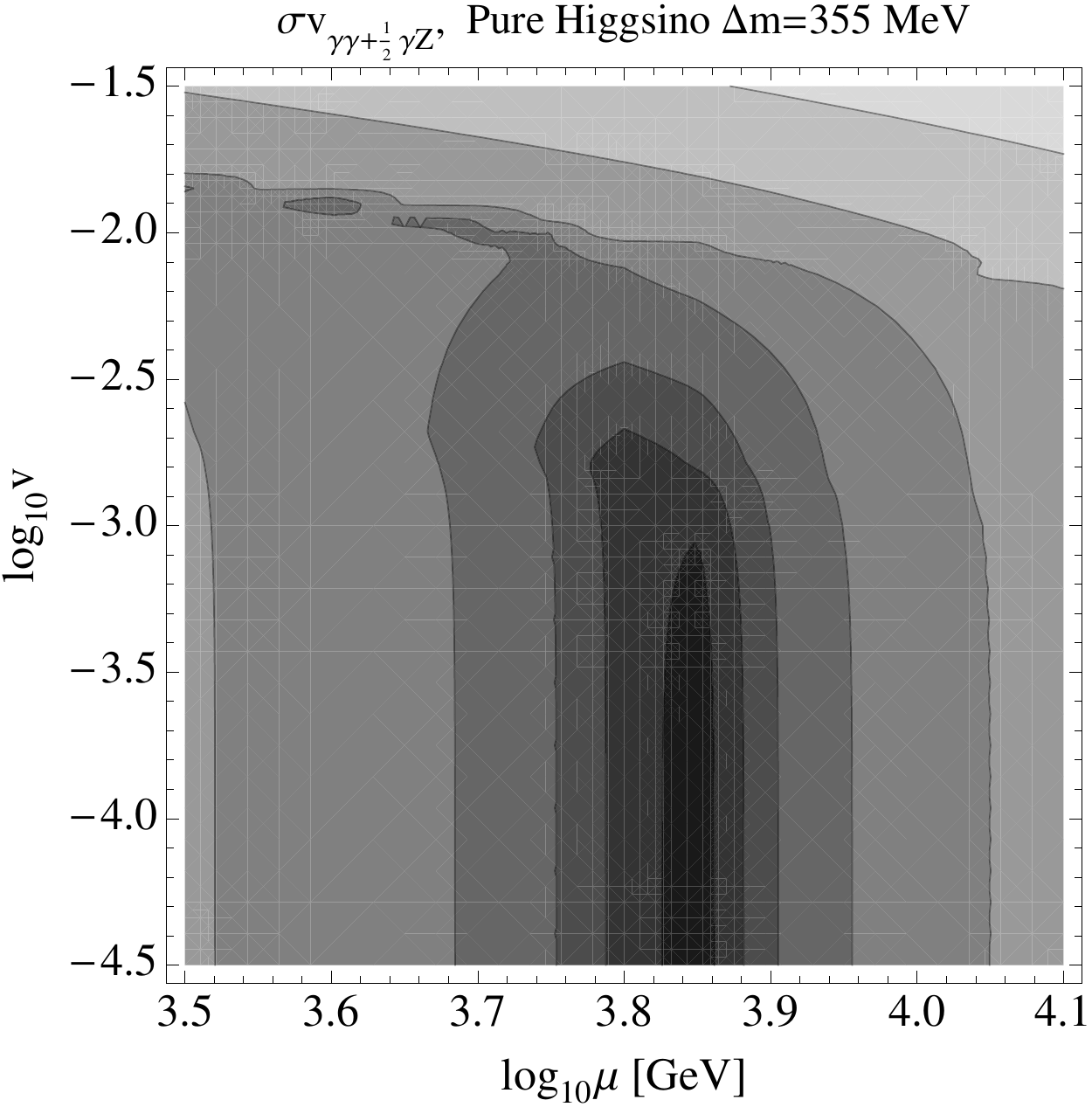}
\hspace{0.2in} \includegraphics[width=0.19\textwidth]{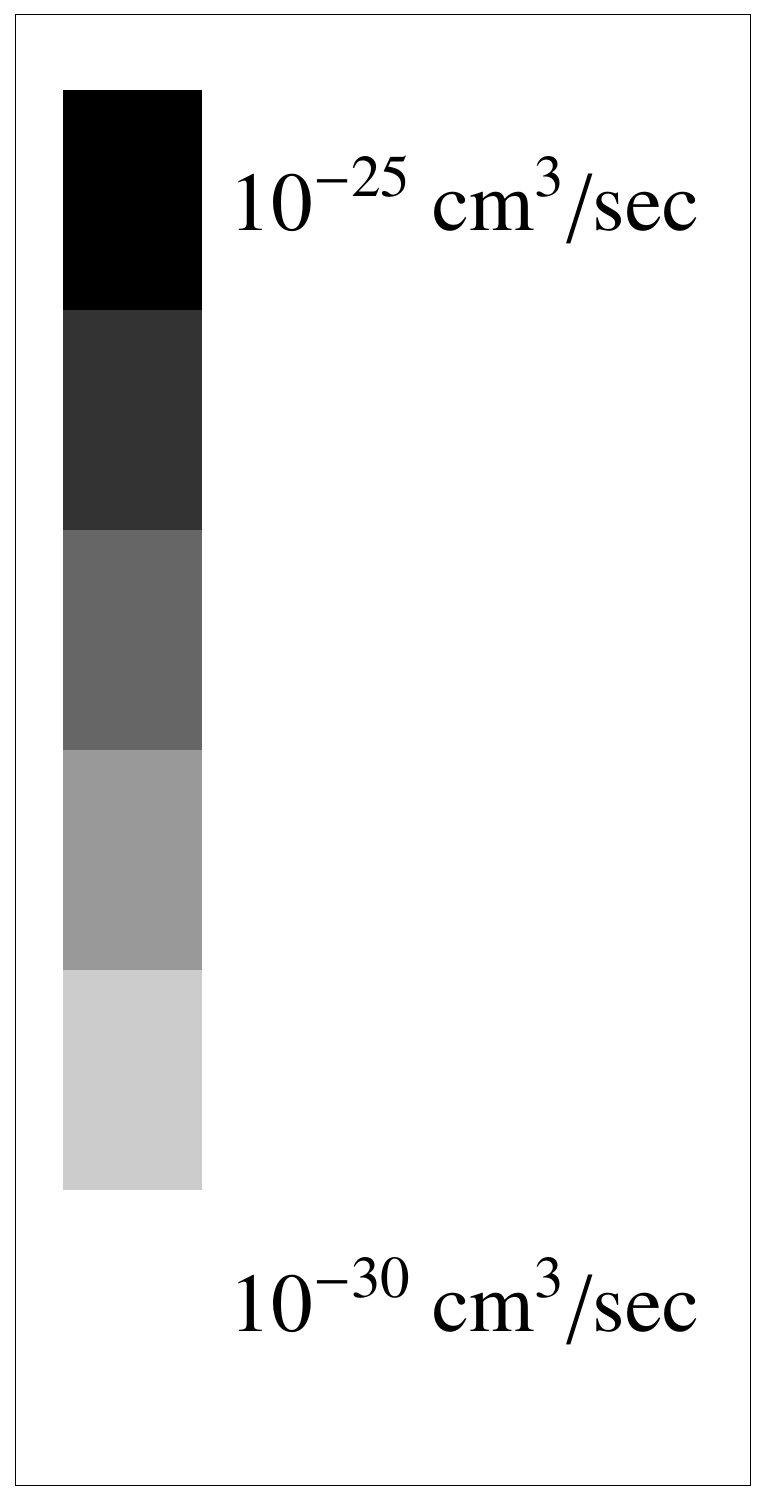}
\caption{Contours of annihilation cross-section into photon-line signals $\sigma v_{\gamma \gamma + \frac{1}{2}\gamma Z}$ for $\Delta m=2$ (top-left), 10 (top-right), 355 (bottom) MeV.
Both enhancements from threshold resonances and RT suppressions are visible; far below the excitation threshold (blue-dashed), SRT effects become velocity-independent.
As $\Delta m$ increases, peaks and dips move to heavier masses and larger velocities, and they become more separated. 
Some irregularities in contours are owing to lack of resolution in parameter scanning.
Very Degenerate Higgsino DM parameters are as in \Fig{fig:mu-m1-masseigen}.
}
\label{fig:contour-S}
\end{figure}

We study two benchmark models with $\Delta m = 2, 10$ MeV presented in \Fig{fig:mu-m1-masseigen}.
For the given $\mu \in \{ 600, 2000 \}$ GeV (and other parameters as described), a unique solution for $M_1$ is found.
As long as gaugino mixtures are small, the exact value of $M_2$ ($\gtrsim |M_1|$) does not matter much in annihilation signals.
It is because leading contributions to annihilations and SRT effects already exist in the pure Higgsino model with vanishing gaugino mixings: 
for example, direct annihilation $\chi^0 \chi^0 \to WW$ and SRT effect $\chi^0 \chi^0 \to \chi_1^+ \chi^-$ can be mediated by the Higgsino-Higgsino-$W$ interaction without need for any gaugino mixtures.
Thus, we set $M_2=10$ TeV (and $t_\beta =1.8$) in this section.

In \Fig{fig:contour-S}, we show contours of annihilation cross-section into photon-line signals, $\sigma v_{\gamma \gamma + \tfrac{1}{2}\gamma Z} \equiv \sigma v_{\gamma \gamma} + \frac{1}{2}\sigma v_{\gamma Z}$, for the benchmark models with $\Delta m = 2, 10$ MeV and the usual pure Higgsino model with $\Delta m =355$ MeV for comparison.
Similar features exist in photon-continuum signals from $\sigma v_{WW+ZZ} \equiv \sigma v_{WW} + \sigma v_{ZZ}$, and similar discussions apply.

Two types of enhancements are observed, most clearly from the $\Delta m=2$ MeV result.
First, a series of threshold zero-energy resonances form just below the excitation threshold of $\chi^0 \chi^0 \to \chi^+ \chi^-$ with $\tfrac{1}{2} \mu v^2 \simeq \Delta m$ (blue-dashed line)~\cite{Slatyer:2009vg, Beneke:2014hja, MarchRussell:2008tu}, depicted as diagonal bands of enhancement.
Photon exchanges between chargino pairs are responsible for the series of closely-located resonances, but not all of them are captured and shown in the figure; see Ref.~\cite{Beneke:2014hja} for  demonstration of many closely-located threshold resonances.
Well below the threshold, SRT effects are independent on velocity as the $W$-boson exchange in $\chi^0 \chi^0 \to \chi^+ \chi^-$ becomes governed by the $W$-mass rather than DM momentum~\cite{Hisano:2003ec, Hisano:2004ds, Beneke:2014hja}, depicted as vertical regions of enhancements. 
The SRT effect saturates at finite enhancement in the $v\to0$ limit because of the finite-ranged $W$-exchange Yukawa potential.

As $\Delta m$ increases, the excitation $\chi^0 \chi^0 \to \chi^+ \chi^-$ becomes harder and the attractive potential becomes effectively shallower~\cite{Hisano:2004ds}.
A heavier DM with a smaller Bohr radius can compensate this trend and can form zero-energy bound states.
Thus, the larger $\Delta m$, the heavier Higgsino Sommerfeld peaks.
From  $\mu \sim 1.1$ TeV for $\Delta m=2$ MeV, the Sommerfeld peak moves to a heavier $\mu \sim 1.3$ TeV for $\Delta m=10$ MeV and to much heavier $\mu \sim 7$ TeV for the pure Higgsino with $\Delta m=355$ MeV.
Moreover, the threshold velocity becomes higher with larger $\Delta m$, making the SRT effects saturate at higher velocities. All such behaviors are clearly shown in \Fig{fig:contour-S}.

Another remarkable is that RT dips are formed near Sommerfeld peaks~\cite{Chun:2012yt, Chun:2015mka, Cirelli:2015bda, Garcia-Cely:2015dda} both in the excitation threshold and in the small-velocity saturation regimes. 
RT dips are located at slightly heavier Higgsino masses and/or larger velocities. 
As $\Delta m$ increases, dips and peaks become more separated in $\mu$ and $v$.

\subsection{Annihilations at GC and DG}

We calculate annihilation cross-sections at GC and DG, main candidate sites for DM indirect detection.
GC is expected to support huge DM density but also plenty of contaminations from baryons, whereas DG is a very clean DM source in spite of smaller DM density.
In addition, velocity dispersions are order of magnitude different, often further differentiating annihilation signals at DG and GC.

We convolute the annihilation cross-section calculated in the previous subsection with Maxwell-Boltzmann velocity distributions for GC and DG~\cite{Feng:2010zp, Essig:2010em, Chun:2012yt, Cannoni:2013bza}
\beq
\langle \sigma v \rangle \= \int dv \, (\sigma v) \cdot \sqrt{\frac{8}{\pi}} \frac{4v^2}{v_0^2} \cdot \exp\left[ - \frac{4v^2}{2v_0^2} \right]\,,
\eeq
which we write in terms of DM velocity $v=v_{\rm DM}$ (in accordance with the SRT calculation in the previous subsection) instead of relative velocity $v_{\rm rel} = 2 v_{\rm DM}$.
The velocity dispersions are chosen to be $v_0 = 210$ km/s for GC and 20 km/s for DG.
Most relevant velocity ranges are $\log_{10} v = -3.2 \sim -3.5$ for GC and $\log_{10} v = -4.1 \sim -4.6$ for DG.

\begin{figure}[t] \centering
\includegraphics[width=0.49\textwidth]{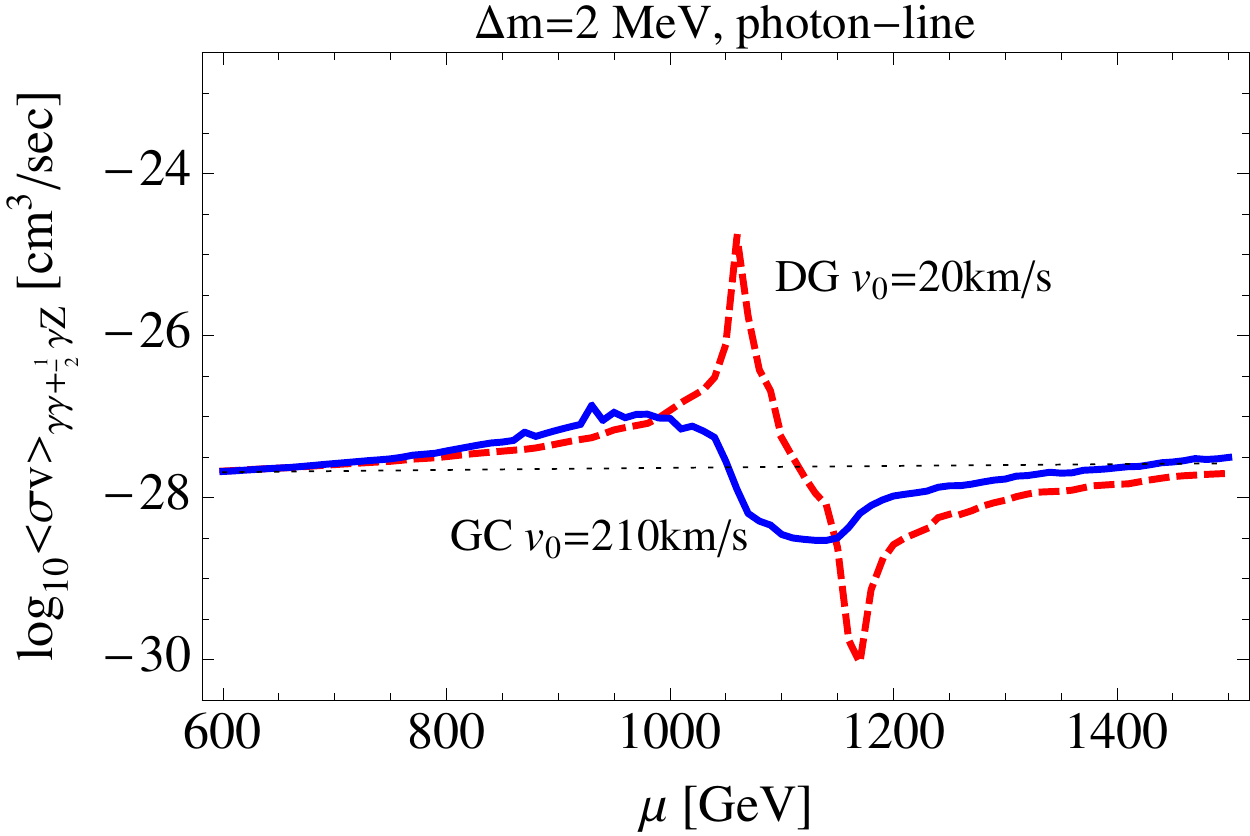}
\includegraphics[width=0.49\textwidth]{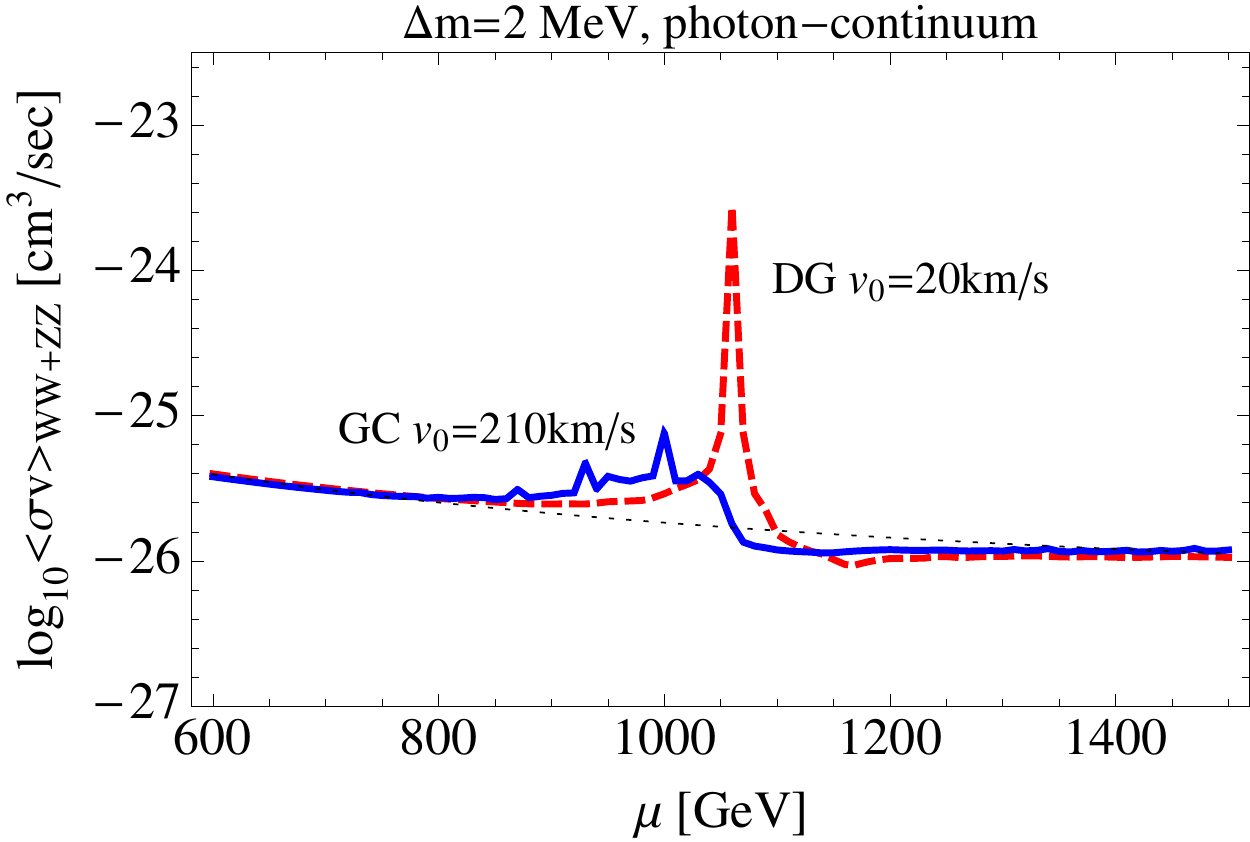}
\includegraphics[width=0.49\textwidth]{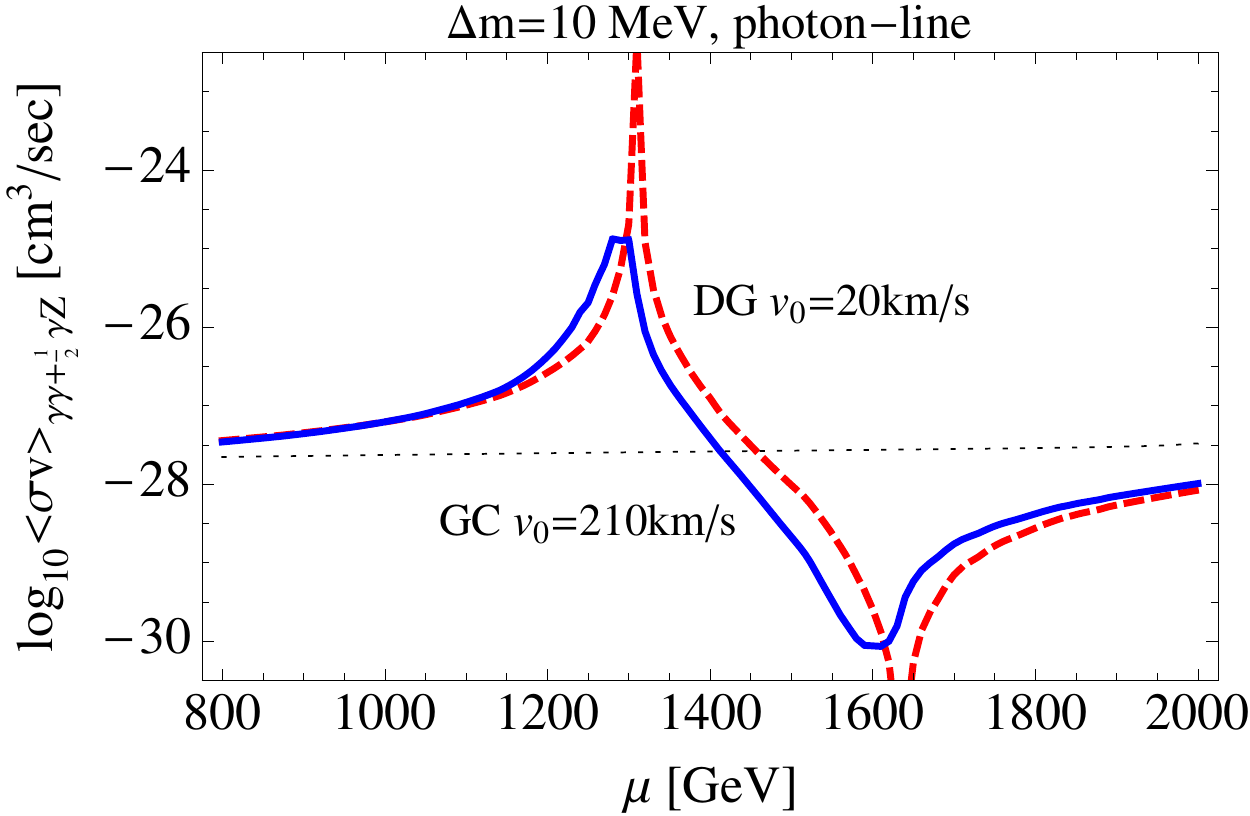}
\includegraphics[width=0.49\textwidth]{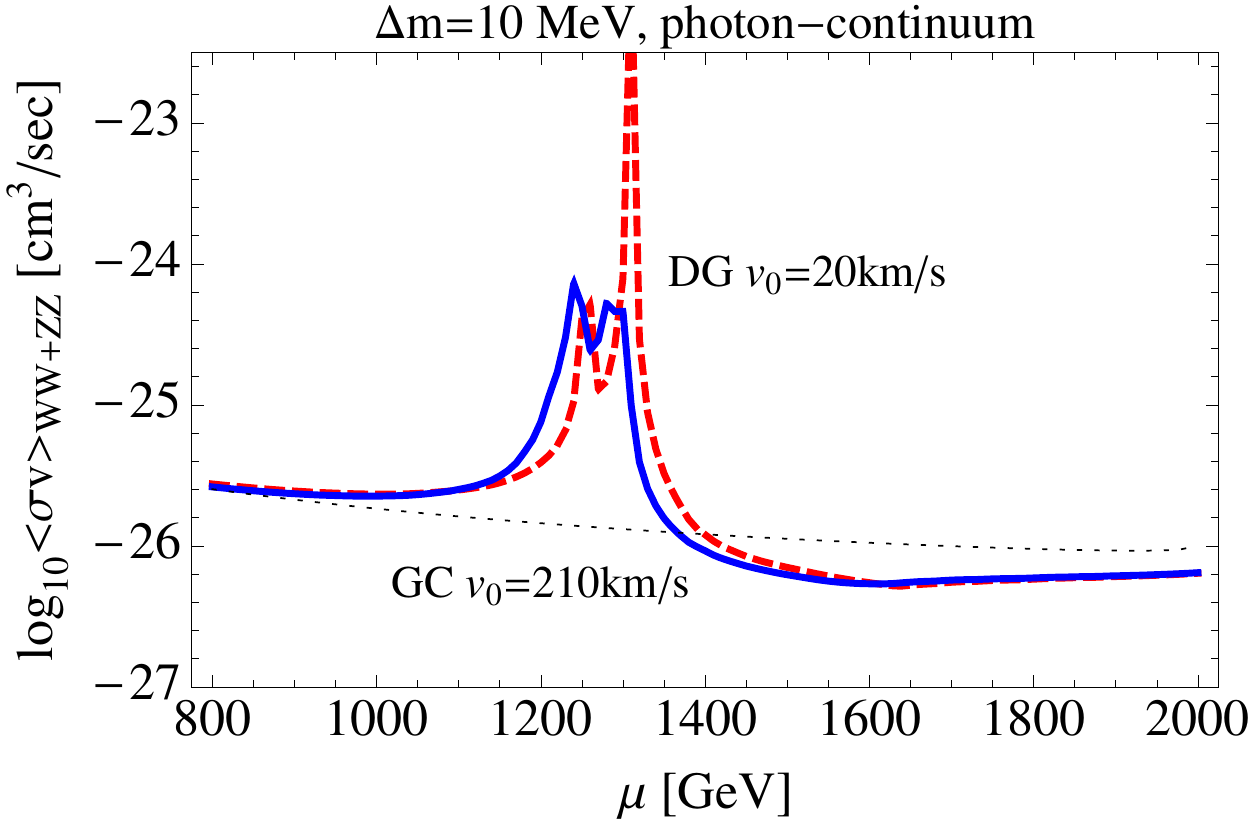}
\caption{Annihilation cross-sections convoluted with Maxwell-Boltzmann velocity distributions.
Velocity dispersions for GC (blue-solid) and DG (red-dashed) are $v_0 = 210$ and 20 km/s. 
Panels are for $\Delta m=2$ (top), 10 MeV (bottom) and photon-line cross-section $\sigma v_{\gamma \gamma+ \frac{1}{2}\gamma Z}$ (left), photon-continuum cross-section $\sigma v_{WW+ZZ}$ (right).
For comparison, perturbative results are also shown (dotted). Some irregularities are owing to lack of resolution in parameter scanning.}
\label{fig:ann-crx}
\end{figure}
\begin{figure}[t] \centering
\includegraphics[width=0.49\textwidth]{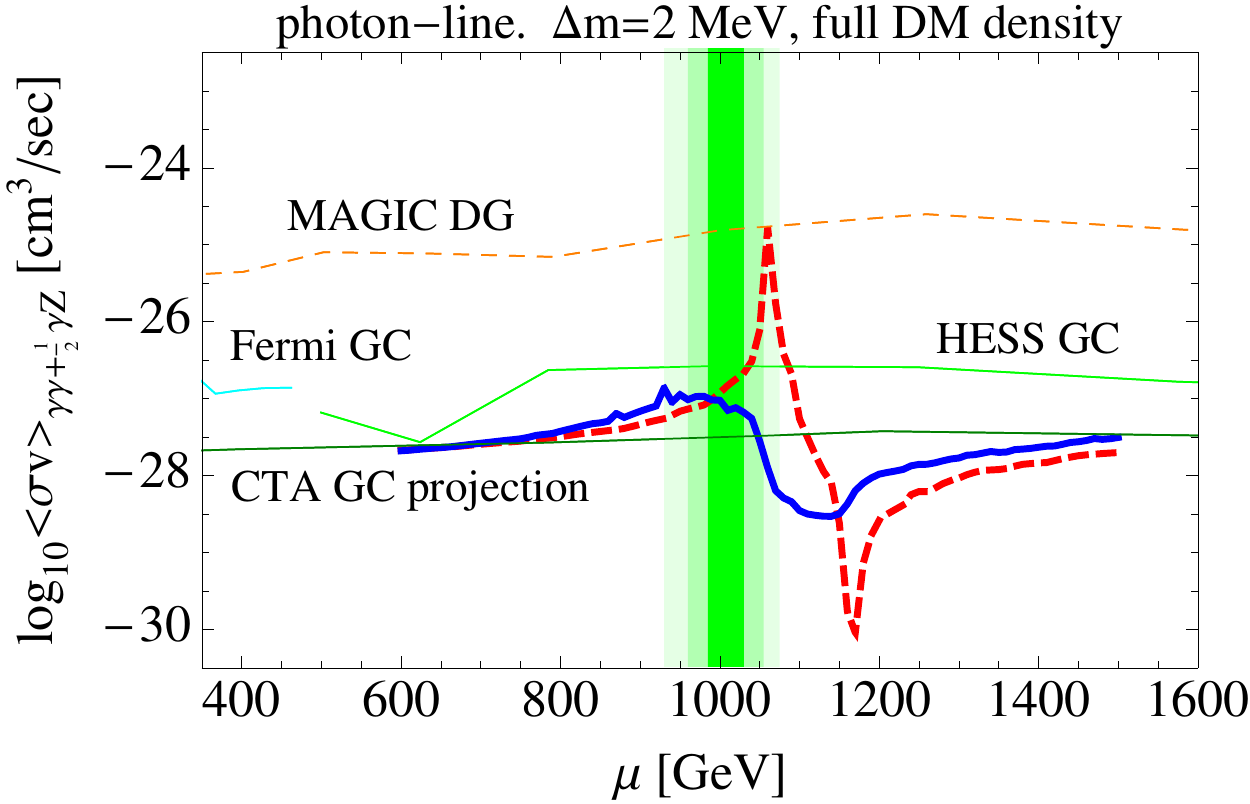}
\includegraphics[width=0.49\textwidth]{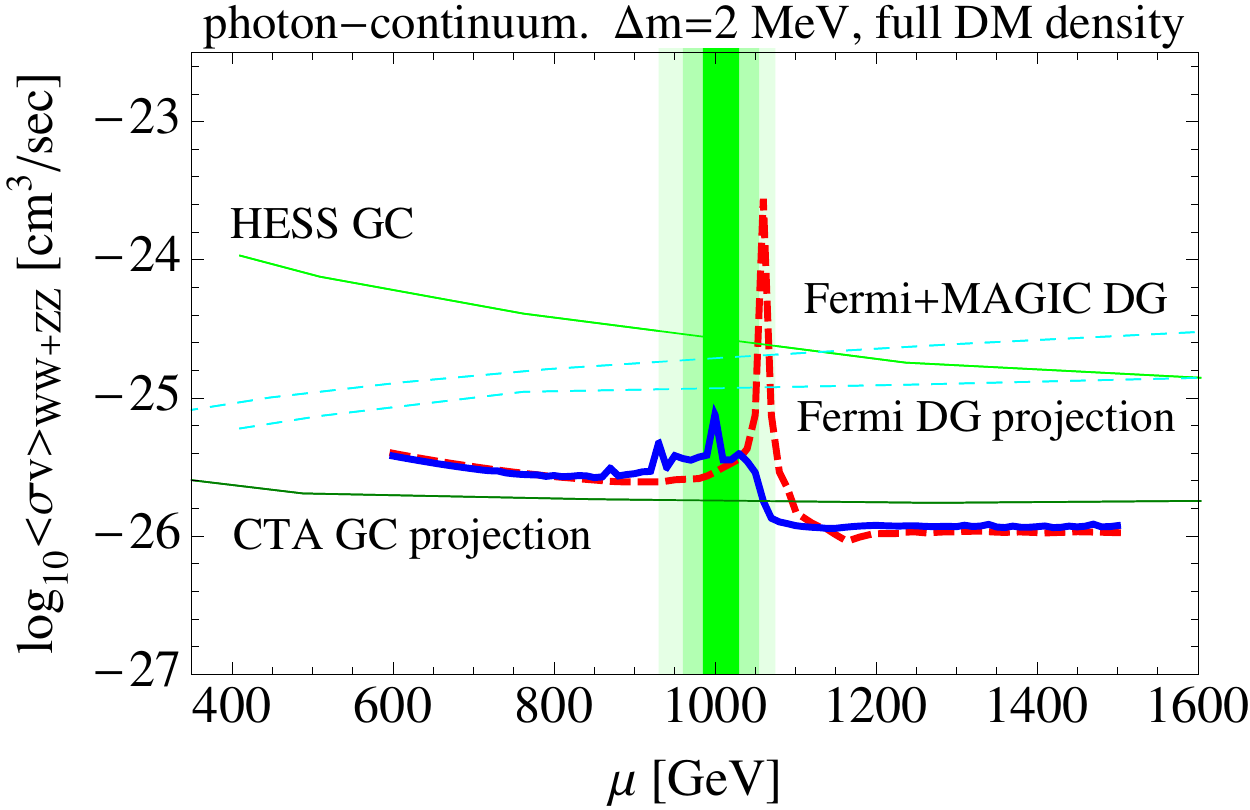}
\includegraphics[width=0.49\textwidth]{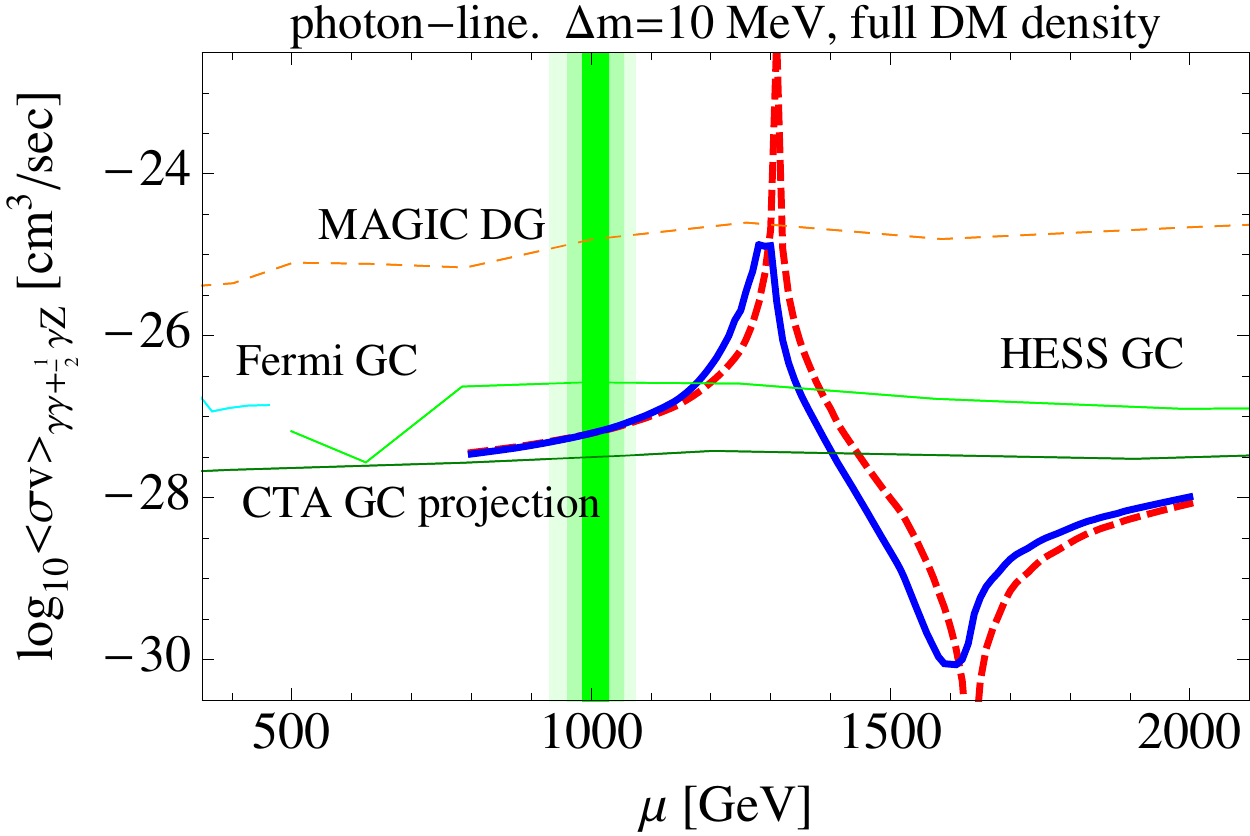}
\includegraphics[width=0.49\textwidth]{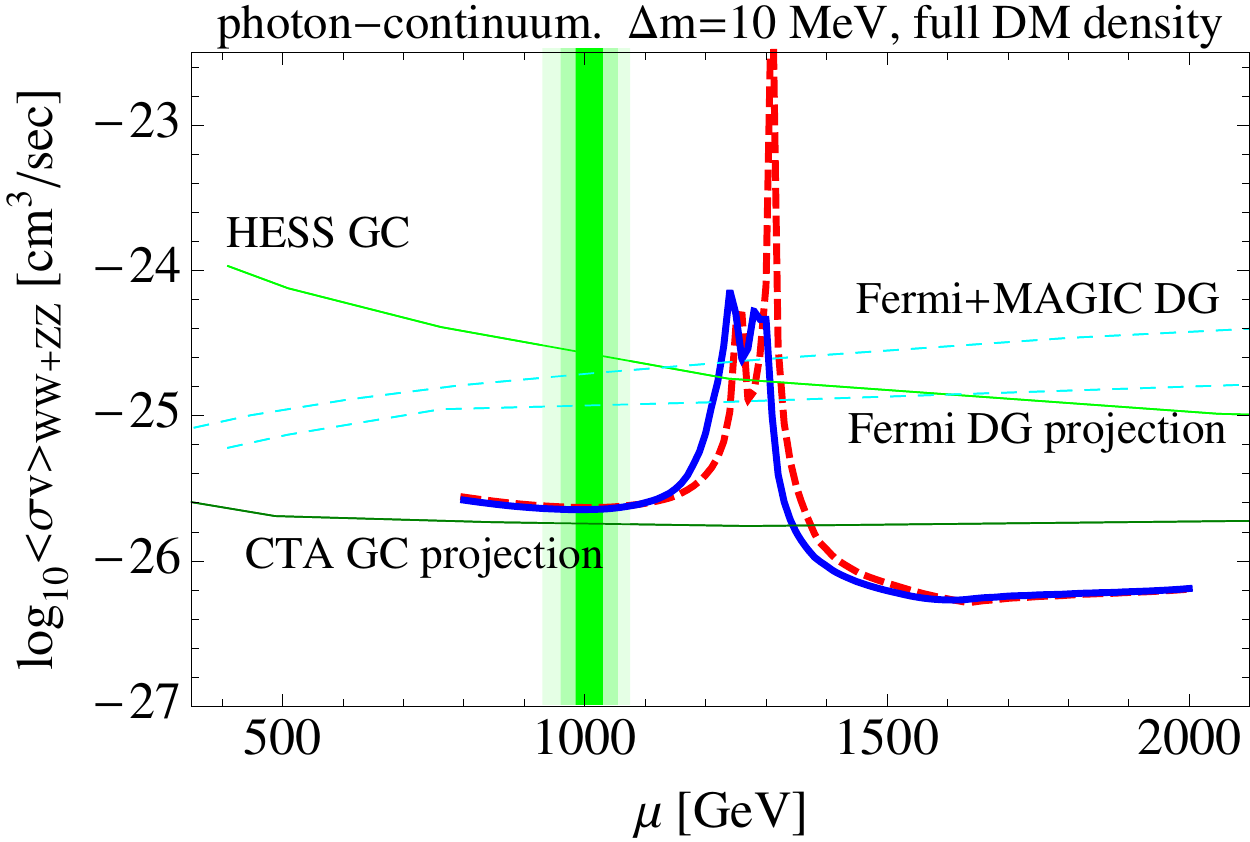}
\caption{Latest constraints from Fermi-LAT (cyan), HESS (light green), MAGIC (orange) and projections from Fermi-LAT (cyan) and CTA (dark green); more details in text.
Panels are for $\Delta m=2$ (top) and 10 MeV (bottom), and photon-line (left) and photon-continuum (right).
Solid lines are for GC and dashed for DG.
Full DM relic density is assumed for all masses; for reference, usual thermal Higgsino DM mass range is shown as green bands.}
\label{fig:ind-const}
\end{figure}

Resulting velocity-convoluted annihilation cross-sections at GC and DG are shown in \Fig{fig:ann-crx}.
Sommerfeld enhancements and RT suppressions are both clearly observed near 1 TeV Higgsino.
Near Sommerfeld peaks and RT dips, annihilation cross-sections at GC and DG are different in general.
The difference is larger for the $\Delta m=2$ MeV case because SRT effects saturate at lower velocity.
Meanwhile, overall enhancements and suppressions are larger for the $\Delta m=10$ MeV case because peaks and dips are more separated in $\mu$ and $v$ so that they lead to less cancellation in velocity convolution.
We also comment that GC cross-sections are not as sharp as DG ones in the figure because we had to average over very closely-separated peaks and dips appearing just below the excitation threshold (where GC signal is most sensitive too) and not all well captured in our parameter scanning.

Another remarkable feature in \Fig{fig:ann-crx} is that, owing to RT dips, DG annihilation cross-section can be smaller than that of GC.
It is a counter-example to the typical result that DG annihilation cross-section is similar or larger than GC annihilation because DM velocity dispersion is smaller.
The existence of RT dips is (accidentally) more clear in the photon-line signal than in the photon-continuum signal; as RT dips are produced from cancellations between various contributions (not necessarily related to resonances), their appearance and strength can depend on annihilation channels.

The exact peak heights shown in the figure may be subject to uncertainties; 
our parameter scanning resolution very close to peak centers is limited and perturbative corrections that may become important in this regime are not added.
The perturbative corrections are most important when unitarity is broken by unphysically enhanced cross-section~\cite{Blum:2016nrz}. 
However, our annihilation cross-sections are well below the unitarity bound $\sigma v \leq 4 \pi/(\mu^2 v) \simeq 10^{-20} \times \left(\frac{1\, {\rm TeV}}{\mu} \right)^2 \left( \frac{10^{-2}}{v} \right) {\rm cm}^3 / {\rm sec}$; 
and indeed, the regularizing velocity $v_c \sim 10^{-6}$~\cite{Blum:2016nrz} is much smaller than our saturation velocity. Also, our scanning resolution is good enough just away from peak centers. Thus, we do not attempt to further improve peak height calculation.

\medskip

In \Fig{fig:ind-const}, we finally overlay the latest constraints and some projection limits of indirect detections.
Datasets presented include: HESS 2013~\cite{Abramowski:2013ax} and Fermi-LAT 2015~\cite{Ackermann:2015lka} for photon-line from GC, MAGIC 2013~\cite{Aleksic:2013xea} for photon-line from DG, Fermi-LAT+MAGIC combination~\cite{Ahnen:2016qkx} for photon-continuum from DG, and HESS 254h~\cite{Lefranc:2015vza, Abazajian:2011ak, Abramowski:2011hc} for photon-continuum from GC.
Projection studies include: CTA 5h~\cite{Bergstrom:2012vd, Ibarra:2015tya, Chun:2015mka} for photon-line from GC, CTA 500h~\cite{Carr:2015hta, Lefranc:2015pza} for photon-continuum from GC, and Fermi-LAT 15 years for photon-continuum from 16 DGs~\cite{Charles:2016pgz}. Current and future DES constraints from DG photon-continuum~\cite{Drlica-Wagner:2015xua} are similar or weaker than the results shown, so we do not show them. Full DM relic density is assumed for all Higgsino masses in interpreting the constraints.

Currently, Sommerfeld peaks in both $\Delta m=2, 10$ MeV models are constrained by DG searches.
Also, GC searches constrain Sommerfeld peaks of the $\Delta m=10$ MeV case, while smaller peaks of the $\Delta m=2$ MeV are not yet constrained by GC searches. In the future, a large part of Sommerfeld enhanced parameter space can be probed by CTA GC  and Fermi DG searches. On the other hand, RT dips in photon-line signals are well below future sensitivities although RT dips in photon-continuum signals are less significant and only midly below the CTA GC projection.

For reference, we also show as green bands the mass range where the thermal Higgsino DM with $\Delta m=355$ MeV can explain the full DM relic density.
Although SRT effects on the Very Degenerate Higgsino model can alter the relic density somewhat, the pure Higgsino result is still a useful guide as SRT effects on relic density may not be so significant.
Not only nearby Sommerfeld peaks and RT dips may cancel each other during thermal history, but also some co-annihilation channels may have opposite SRT effects (as for the pure Higgsino DM~\cite{Beneke:2014hja}) that can also nullify impacts on relic density.
Without dedicated relic density calculation, we are content with assuming the full DM relic density, and in any case our signals can be scaled in proportion to true relic density.

\section{Other Constraints}

\subsection{Direct Detection} \label{sec:dd}

The spin-independent direct detection (SIDD) signal of the nearly degenerate Higgsino DM depends on the mass splitting between the neutral states $\delta m^0$, and the amount of the gaugino mixture.
The neutral mass gap $\delta m^0$ should be larger than ${\cal O}(0.1)$ MeV,
otherwise its inelastic scattering mediated by $Z$ exchange should have been already observed~\cite{Chun:2012yt}.
For sufficiently large $\delta m^0$ as in our study (see \Fig{fig:mu-m1-masseigen}), the elastic scattering rate is controlled by gaugino mixtures (via Higgsino-gaugino-Higgs coupling), that is, the signal vanishes in the pure Higgsino limit.
Therefore, we consider two benchmark values of $M_2 = 10$ and 5 TeV in this subsection, representing the cases with relatively small and large gaugino mixings and SIDD signals.
For each $M_2$ benchmark, the value of $M_1$ is fixed (as a function of other parameters) to obtain the desired $\Delta m=2, 10$ MeV, and thus SIDD rates are determined.

The SIDD cross-section is approximately given by~\cite{Cheung:2012qy}
\beq
\sigma_{\rm SI} \simeq 8\times 10^{-47} \left( \frac{g_{h\chi \chi}}{0.01} \right)^2 \, {\rm cm}^2, \qquad \qquad {\cal L} \, \ni \, g_{h\chi \chi} \, \overline{\chi^0_1} \, \chi^0_1 \, h
\eeq
\beq
g_{h\chi \chi} \= g \Big( N_{12} - t_W N_{11} \Big) \Big( N_{14} s_\beta - N_{13} c_\beta \Big) \, \simeq \, \mp g \frac{m_W}{2 M_2} ( 1 \mp  s_{2\beta} ) \, \Big(\epsilon_K \mp \frac{\mu}{M_1} (1+t_W^2) \Big)\,,
\label{eq:hxx} \eeq
where the sign $\mp$ implies the sign(-$\epsilon_K$) and we assume the Higgs alignment limit.
We obtain $\sigma_{\rm SI} \= (3 \sim5) \times 10^{-48}, \, (4\sim9) \times 10^{-47} \, {\rm cm}^2$ for $M_2 = 10, \, 5$ TeV with the range spanned by $\mu = 600 \sim 1500$ GeV (see \Fig{fig:mu-m1-masseigen} for $M_2=10$ TeV result). The dependence on the $\Delta m $ (indirectly via Bino mixtures) is not significant for $\Delta m \lesssim 10$ MeV.
The former range of $\sigma_{\rm SI}$ with $M_2=10$ TeV is close to the coherent neutrino scattering background floor so that searches will be very difficult in the near future, while the latter range with $M_2=5$ TeV is expected to be probed at future experiments such as DarkSide-G2~\cite{Cushman:2013zza, Aalseth:2015mba} and LZ~\cite{Cushman:2013zza, Akerib:2015cja}.
Although indirect detection signals are sizable for both $M_2$ benchmark values, the absence or existence of detectable SIDD signal still depends on the Wino mixture (hence, the Wino mass) and is not a necessary consequence of the Very Degenerate Higgsino DM.

\subsection{Collider Searches}

With very small mass splitting, the charged Higgsino can be long-lived at LHC experiments.
If it decays outside or outer part of LHC detectors, stable chargino searches apply, that is, characteristic ionization pattern of traversing massive charged particles can be identified.
If it decays in the middle of detectors, disappearing charged track searches apply as soft charged decay products are not efficiently reconstructed.

For $\Delta m$ much smaller than the pion mass, the dominant chargino decay mode is $\chi^+ \to e^+ \nu_e \chi^0$~\cite{Thomas:1998wy, Chen:1996ap, Martin:2000eq, Jung:2014bda}:
\beq
\Gamma( \chi^+ \to e^+ \nu_e \chi^0 ) \= \frac{G_F^2}{30\pi^3} (\Delta m)^5 \, \sqrt{1-\left(\frac{m_e}{\Delta m}\right)^2 } \, P( m_e / \Delta m)
\label{eq:chargino-width} \eeq
with the function $P(x)$ given in Ref.~\cite{Thomas:1998wy}.
For $\Delta m \sim {\cal O}(1-10)$ MeV, the decay length is very long, $c\tau \sim 10^{7} - 10^{12}$ m (equivalently $\tau \sim 10^{-1} - 10^4$ sec), so that almost all charginos traverse LHC detectors and thus only stable chargino searches apply.

Reinterpreting the CMS 8 TeV constraints on the stable charged pure Wino~\cite{Khachatryan:2015lla}, we obtain the constraint $\mu \gtrsim 400-600$ GeV for $\Delta m$ much smaller than the pion mass.
The uncertainty range quoted is partly owing to our lack of knowledge of $r_{\rm min}$, the minimum decay length of the chargino for the stable chargino search to be applied; it is needed because CMS considered the range of charged Wino decay length $c\tau = {\cal O}(0.1-10)$ m where only a fraction of charged Winos traverse detectors and become stable charginos.
From the CMS acceptance curve in Ref.~\cite{Khachatryan:2015lla}, we choose to vary $r_{\rm min} = c\tau_{\rm min} \simeq 1.5 - 6$ m ($\tau_{\rm min} = 5-20$ ns) to obtain the constraint and uncertainty.

We conclude that the $\sim$ 1 TeV Very Degenerate Higgsino DM is currently allowed, but future LHC searches of stable charginos will better constrain the model.

\subsection{Cosmological Constraints}

The long-lived charged Higgsino can be cosmologically dangerous. 
The above quoted lifetime in our model $\tau \sim 10^{-1} - 10^4$ sec could endanger the standard bing-bang nucleosynthesis (BBN) prediction.
Although the chargino decay releases only soft leptons not directly affecting BBN, its metastable existence can form a bound state with a helium and can catalyze the $^6$Li production.
The lifetime limit $\tau \lesssim 5000$ sec of such a metastable charged particle~\cite{Pospelov:2006sc} constrains the Higgsino mass splitting to be $\Delta m \gtrsim 1.2$ MeV.\footnote{The limit on stau-neutralino mass splitting, 70 MeV, reported in Ref.~\cite{Pospelov:2006sc} is much stronger because the stau has four-body decays and is thus longer-lived.}
The $(\Delta m)^5$ dependence of the decay width in \Eq{eq:chargino-width} makes the BBN constraints quickly irrelevant to larger $\Delta m$ cases that we focus on.

As the enhancement is saturated at modestly small velocity, early-universe constraints from the era with very small DM velocity such as recombination and DM protohalo formation are not strong.
For example, $\sigma v_{WW} \lesssim 10^{-24} \, {\rm cm}^3/ {\rm sec}$ is generally safe from such considerations (see, e.g., Refs.~\cite{Kamionkowski:2008gj,Galli:2009zc,Slatyer:2009yq}), so that the model is not constrained possibly except for a very small parameter space close to Sommerfeld peaks.

\section{Summary and Discussions}

We have studied the Very Degenerate Higgsino DM model with ${\cal O}(1)$ MeV mass splitting, which is realized by small gaugino mixing and leads to dramatic non-perturbative effects.
Owing to the very small mass splitting, SRT peaks and dips are present at around 1 TeV Higgsino mass, and velocity saturation of SRT effects is postponed to lower velocities $ v/c \sim 10^{-3}$.
As a result, indirect detection signals of $\sim 1$ TeV Higgsino DM can be significantly Sommerfeld-enhanced (to be constrained already or observable in the near future) or even RT-suppressed.
Annihilation cross-sections at GC and DG are different in general: 
either of them can be larger than the other depending on the location of Sommerfeld peaks and RT dips.
Other observable signature is also induced in stable chargino collider searches, which can probe the 1 TeV scale in the future.
However, the rates of direct detection signals depend on the $M_2$ value (the smaller $M_2$, the larger signal) so that $M_2 \sim 5$(10) TeV can(not) produce detectable signals.
Because of various unusual aspects of indirect detection signals at DG and GC, well featured by our two benchmark models of $\Delta m=2$ and 10 MeV, future searches and interpretations on Higgsino DM models shall be carefully done.

The Very Degenerate Higgsino DM also provides an example where ``slight'' gaugino mixing can have unexpectedly big impacts on the observation prospects of the Higgsino DM.
The mixing is slight in the sense that direct detection, whose leading contribution is induced by gaugino mixing, can still be small (for heavy enough Winos).
At the same time, however, phenomenology is unexpectedly interesting because such slight mixing could significantly change indirect detection signal, which is present already in the zero mixing limit so that usually thought not to be so sensitive to small mixing.
In all, \emph{nearly} pure Higgsino DM can have vastly different phenomena and discovery prospects from the pure Higgsino DM, and we hope that more complete studies can be followed.

\acknowledgments
We thank Kyu Jung Bae and Ranjan Laha for discussions on cosmological constraints and Kfir Blum on regulating Sommerfeld peaks.
The work of SJ is supported by the US Department of Energy under contract DE-AC02-76SF00515 and in part by the National Science Foundation under Grant No. NSF PHY11-25915.
JCP is supported by the Basic Science Research Program through the National Research Foundation of Korea (NRF-2013R1A1A2061561, 2016R1C1B2015225). 
SJ thanks KITP for their hospitality during the completion of the work.


\end{document}